# Modeling supply chain compliance response strategies based on AI synthetic data with structural path regression: A Simulation Study of EU 2027 Mandatory Labor Regulations

Wei Meng

Dhurakij Pundit University

Fellow of The Royal Anthropological Institute of Great Britain and Ireland

wei.men@dpu.ac.th

## Abstract

In the context of the new mandatory labor compliance in the European Union (EU), which will be implemented in 2027, supply chain enterprises face stringent working hour management requirements and compliance risks. In order to scientifically predict the enterprises' coping behaviors and performance outcomes under the policy impact, this paper constructs a methodological framework that integrates the AI synthetic data generation mechanism and structural path regression modeling to simulate the enterprises' strategic transition paths under the new regulations. In terms of research methodology, this paper adopts high-quality simulation data generated based on Monte Carlo mechanism and NIST synthetic data standards to construct a structural path analysis model that includes multiple linear regression, logistic regression, mediation effect and moderating effect. The variable system covers 14 indicators such as enterprise working hours, compliance investment, response speed, automation level, policy dependence, etc. The variable set with explanatory power is screened out through exploratory data analysis (EDA) and VIF multicollinearity elimination. The findings show that compliance investment has a significant positive impact on firm survival and its effect is transmitted through the mediating path of the level of intelligence; meanwhile, firms' dependence on the EU market significantly moderates the strength of this mediating effect. It is concluded that AI synthetic data combined with structural path modeling provides an effective tool for high-intensity regulatory simulation, which can provide a quantitative basis for corporate strategic response, policy design and AI-assisted decision-making in the pre-prediction stage lacking real scenario data.

# Chapter I. Introduction

Against the backdrop of the continuous upgrading of global value chain governance, the EU's Compulsory Labor Regulation, which will be formally implemented in 2027, marks the full transformation of cross-border supply chain regulation from moral persuasion to legal compulsion. The regulation stipulates that enterprises exporting to the EU must ensure that their employees in the entire supply chain work no more than eight hours a day, and deploy face-recognition work hour recording systems, accept on-site audits and unannounced spot checks. Any violators will face high costs in terms of product bans, order freezes and even legal action. While this regulation places historic demands on supply chain ethical standards, it also poses serious challenges to the digital transformation, cost control and compliance capabilities of manufacturing companies.

Although the regulations have yet to come into effect, their impact is already generating a great deal of concern globally. Especially for Asian manufacturing and cross-border e-commerce companies, whose competitiveness is based on the "artificial dividend", how to plan compliance paths and evaluate cost structures and performance changes in advance in a highly uncertain institutional environment has become an urgent strategic decision-making issue. Traditional policy evaluation studies rely on real observation data after the implementation of regulations, which is difficult to adapt to "forward-looking simulation" scenarios where regulations have not yet come into effect and there is a lack of empirical data.

Existing literature has conducted many explorations on AI synthetic data, causal modeling, path regression, and corporate strategic response, but there are still theoretical and methodological gaps in the following key areas: first, the lack of a synthetic data modeling methodology chain that can be adapted to high-intensity regulatory scenarios, second, the mediator-regulator composite paths involved in supply chain strategic behavior have not yet been incorporated into a unified analytical framework, and third, the structural causal mechanism between firms' strategic choices and performance outputs still lacks an AI-driven high-confidence predictive model. The structural causal mechanism still lacks AI-driven highly credible predictive models. Especially before the regulations officially come into effect, building modeling mechanisms that can be used for response strategy evolution analysis and policy impact prediction has become a core challenge for AI policy simulation and decision modeling.

Based on this, this paper proposes a supply chain policy response strategy modeling framework that integrates AI synthetic data generation mechanism and structural path regression modeling. Specifically, the study constructs a structural path system with multiple regression models, logistic regression models, mediation effects and moderating effects to simulate compliance investment, automation transformation, response speed and performance under the pressure of EU regulations. The methodology combines NIST synthetic data standards, VIF variable screening and path effect significance testing to form a set of policy simulation pathways applicable to scenarios without real samples.

The objectives of this study include: (1) to establish a set of AI modeling system with structural consistency of variables and explainable causal paths in the absence of real data; (2) to identify the multi-path response mechanism of "Survival-Cost-Order" of firms under the influence of compliance investment, market structure and governmental support; (3) to assess the effectiveness of firms' strategies under the strong regulatory policies of the European Union and quantify the degree of influence and adjustment conditions. "(2) Identify the multi-path response mechanism of enterprises under the influence of compliance investment and market structure and government support; (3) Evaluate the effectiveness of enterprises' strategies under the EU's strong regulatory policies, and quantify the degree of influence and adjustment conditions.

This paper focuses on the following key research questions:

Question 1 . How can AI techniques be utilized to construct highly credible datasets of simulated enterprises without historical sample data?

Question 2. Does compliance investment indirectly affect firm survival through mediating paths (e.g., automation level enhancement)? What is the significance of its path?

Question 3. Does firms' dependence on the EU market or local government support moderate the strength and direction of this mediating path?

Question 4. What combinations of variables improve order stability or market survivability while controlling for cost risk?

The contribution of this study is to fill the gap in the application of AI synthetic data in supply chain policy modeling, provide an AI modeling paradigm that can be generalized to ESG evaluation, antitrust gaming, and trade barrier previewing scenarios, and provide a data-supported prospective decision-making tool for governments and enterprises to cope with high-strength regulatory scenarios.

# Chapter II. Literature review

**2.1 Application of Artificial Intelligence in Supply Chain Compliance Modeling**

As supply chains have risen in complexity, artificial intelligence (AI) technologies are being used in supply chain management with greater frequency. Studies have observed that AI technologies are advantageous for enhancing the efficiency, flexibility, as well as sustainability of supply chains. For instance, Samuels et al. (2024) pointed out the positive role of AI in supply chain operations, primarily in enhancing operational efficiency, strategic innovation, as well as sustainability[1]. AI-based predictive analytics are also stated to be central in supply chain optimization, driving operational efficiency, cost savings, as well as customer satisfaction[2].

However, its use in compliance modeling is not without challenges despite its much-hyped potential in supply chain management. For instance, data quality, complexity of models, as well as costly implementation, constrain the use of AI in forecasting demands[3]. Moreover, incorporation of AI-enabled technologies must overcome technological as well as cultural hindrances in an organization for its full potential in supply chain compliance modeling.

**2.2 The role of synthetic data in AI modeling**

Against the tide of increasing concerns regarding data privacy and protection, synthetic data is gaining traction as an alternate solution over real data. Studies have established that synthetic data is capable of delivering quality data for AI model training as well as testing without posing risks of exposing sensitive information. For instance, Godbole (2025) explores the use of synthetic data for developing strong AI models in regulated sectors, citing its benefits in terms of diversity of data and compliance[4].

In addition, Belgodere et al. (2023) propose a comprehensive auditing framework for assessing the trustworthiness of synthetic datasets and AI models, emphasizing the need to weigh trustworthiness controls when generating synthetic data[5].

However, the use of synthetic data also faces challenges. For example, Arthur et al. (2023) point out that there are generation, infrastructure, governance, compliance, and adoption challenges to deploying privacy-preserving synthetic data in the enterprise[6]. In addition, the quality and representativeness of synthetic data may affect the performance and reliability of AI models.

**2.3 Application of structural path modeling to supply chain compliance**

Structural Path Modeling (SPM), a statistical method used to analyze complex relationships between variables, has important applications in supply chain compliance research. Researchers have noted that Structural Path Modeling can help identify and quantify the critical factors and paths that affect supply chain compliance. As an example, Samuels et al. (2024) emphasize the importance of structural path modeling in analyzing the role of AI technologies in supply chain management, with particular emphasis in the context of the shift from Industry 4.0 to Industry 6.0[1]. In addition, Wamba et al. (2022) examined AI technology integration in supply chain management, with emphasis on the significance of employing structural path modeling in analyzing the role of AI technologies in organizational performance[7].

However, the application of structural path modeling also faces challenges. For example, the complexity of the model and the availability of data may limit its application in practice. In addition, structural path modeling requires high-quality data support, which may be difficult to obtain in supply chain compliance studies.

**2.4 Research Gaps and Future Directions**

Although existing studies have made progress in AI technology, synthetic data and structural path modeling, there are still research gaps in the area of supply chain compliance modeling. To start with, current literature is primarily concerned with supply chain management through the use of AI technology and fails to deeply investigate its role in compliance modeling. Second, while synthetic data use offers benefits in terms of security as well as data privacy, its use in supply chain compliance modeling is not adequately researched.Finally, structural path modeling has potential in analyzing supply chain compliance factors, but its practical application still faces challenges such as data and model complexity.

Therefore, future research should focus on the following directions:

1. explore the specific application of AI technologies in supply chain compliance modeling, especially in predicting and optimizing compliance strategies.

2. investigate the application of synthetic data in supply chain compliance modeling and assess its effectiveness in terms of data privacy protection and model performance.

3. develop and optimize structural path modeling approaches to better analyze and understand the complex relationships among supply chain compliance factors.

4. integrate AI techniques, synthetic data, and structural path modeling to build a comprehensive supply chain compliance modeling framework to support enterprise decision-making and strategizing on compliance.

# Chapter III. Research methodology

This study constructs a framework for modeling supply chain compliance response strategies that incorporates AI synthetic data and structural path modeling. The methodological path is shown below::

Policy Scenarios → Variable Systems → Synthetic Data Generation → EDA → Multivariate Modeling → Mediation and Moderation Tests → Strategy Evaluation

This chapter will systematically describe the variable design basis, the modeling logic, the technical path of data generation and its statistical soundness and AI controllability.

**3.1 Variable system construction**

Based on the regulatory elements of the EU 2027 mandatory labor regulation for supply chains and the real-world context of firms' transformation paths, this paper constructs a structural path model covering four types of variables:

**Table 1: Variable Coverage in Structural Path Model (Four Categories)**

| Variable Type | Code | Name | Description |
|---|---|---|---|
| independent variable （X） | X1–X5 | Labor hours, automation levels, compliance investments, responsiveness, etc. | Describe the behavioral attributes of the business |
| intermediary variable | M1–M2 | Increased automation level, cost control | Characterization of path transit mechanisms |

| | | capability | |
|---|---|---|---|
| moderator variable (MOD) | MOD1–MOD3 | Market dependence, industry type, government support | Adjustment of inter-variable path strength |
| implicit variable (Y) | Y1–Y3 | Business survival status, cost growth rate, order change rate | Indicators for evaluating the effectiveness of the final strategy |

All variables were extracted through theoretical constructs, industry literature and interview simulation scenarios, and statistically analyzed by simulated data to verify testability and logical consistency.

**3.2 Model Structure and Path Assumptions**

In this paper, we use a structural path modeling strategy to divide the modeling logic into:

Direct path (X → Y): does compliance behavior directly affect firm performance?

Mediating path (X → M → Y): does it exert its influence indirectly through mechanisms such as intelligence or cost capability?

Moderating path (MOD × M → Y): does it modulate the strength of the mediating path due to market structure, local support, etc.?

Take "compliance investment (X3) → intelligence level (M1) → enterprise survival (Y1)" as the main path, and construct a regulation-mediation linkage model of "X3 × MOD1 → M1 → Y1".

**3.3 AI synthetic data generation methods**

Since the EU regulations have not yet landed and the real data of enterprises are not yet available, this paper adopts the AI synthetic data simulation mechanism for sample construction with the following methods:

1. variable distribution modeling: each variable is set to a reasonable normal or symmetric distribution (e.g., $N(\mu, \sigma^2)$), and the mean and standard deviation refer to the historical enterprise data or expert estimation;

2. path logic embedding: construct mediating and dependent variables through weighting function to simulate real-world causal direction;

3. noise control: introducing Gaussian noise ($\varepsilon \sim N(0, \sigma)$) with mean value 0 to make the model moderately volatile;

4. sample size and feature engineering: set the sample size to 150, normalize the variables, and apply One-Hot coding to the categorical variables;

5. quality control mechanism: apply the NIST "Synthetic Data Framework" standard to control the simulation quality, to ensure the consistency of data structure, the reasonableness of statistical distribution, regressibility and logical interpretability.

**3.4 Exploratory Analysis and Modeling Steps**

After data preparation, the following multi-stage modeling process was used:

**Table 2. Modeling Process Overview**

| Steps | Element | Methodology and rationale |
|---|---|---|
| EDA | Descriptive statistics, distribution of variables, correlation tests | Mean, variance, histogram, heat map |
| Multicollinearity rejection | Variance Inflation Factor (VIF) | Retention of VIF ≤ 5 variables |
| Multiple linear regression (MLR) | Modeling for Y2/Y3 | OLS model with stepwise regression optimization |
| Logistic regression | Modeling for Y1 | Binary Logit Model |
| Mediation effect test | Baron & Kenny 3-step method + Bootstrap confidence test | Path Significance + Hypothesis Testing |

| Moderating effects test | Interaction term modeling + cross term significance test | X*MOD → Y path fitting |

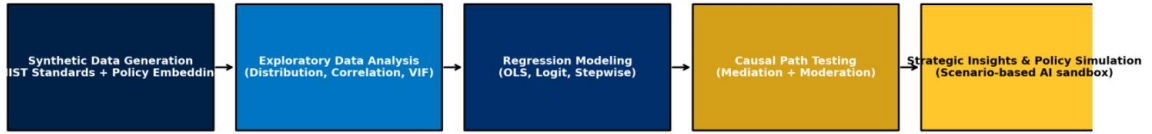

Figure 1：AI Modeling Pipeline

As shown in Figure 1, the overall modeling process in this study follows five interconnected AI-assisted modeling phases, systematically realizing a closed-loop path from data generation to strategic insights. The first stage is synthetic data generation, which constructs an enterprise simulation sample with strong structural consistency and reasonable distribution of variables by integrating NIST structural standards and regulatory scenario embedding mechanism, which is used as a substitute for real observation data. The second stage is exploratory data analysis (EDA), which covers variable distribution test, multicollinearity diagnosis (VIF), and correlation thermal analysis, laying the foundation for model construction. The third stage is regression model construction, using OLS and Logit regression methods, combined with stepwise regression strategy to screen the best explanatory variables and portray the multivariate causal mechanism. The fourth stage is the path effect test, which identifies the indirect and conditional effects of compliance variables on performance output through the analysis of mediating and moderating effects. In the end, combining the regression results and path significance, we output strategic insights and policy simulation recommendations, and construct an AI simulation sandbox based on contextual settings, which can be used to assist in policy scenario derivation and response strategy optimization.

**3.5 Summary of methodology**

The modeling framework constructed in this chapter has the following features:

AI techniques are used to overcome the modeling dilemma of missing pre-policy data; Multi-path structural analysis can explain direct, indirect and moderating effects simultaneously;The modeling logic is scalable and applicable to more scenarios such as green supply chain and antitrust simulation.

# Chapter IV  Modeling and Analysis of Results

Based on the aforementioned AI synthetic dataset and structural path variable system, this chapter develops modeling and empirical analysis around four core research questions. The study uses standardized data, multiple regression, logistic regression, and mediated adjustment modeling to quantify the causal mechanism of the constructed simulation samples and to verify the feasibility and strategy sensitivity of AI-assisted policy prediction modeling.

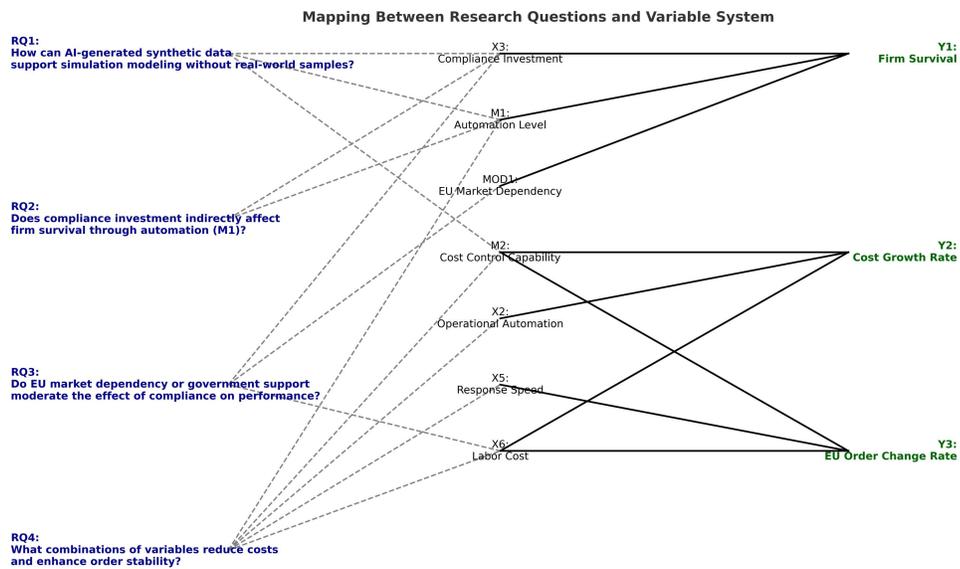

Figure 2：Q ↔ Variable Map

As shown in Figure 2, the Research Question - Variable System Mapping Diagram constructed in this study clearly demonstrates the causal logical relationships and structural nesting between the four core research questions (RQ1-RQ4) and the key modeling variables. RQ1-RQ4) and key modeling variables are clearly shown in a causal logic and structural nesting. On the left side of the diagram, the four research questions are listed sequentially, representing the theoretical main lines of the research design; the nodes in the middle correspond to the selected core independent variables (e.g., compliance investment X3, response speed X5), mediator variables (automation level M1, cost control capability M2), and moderator variables (EU market dependence MOD1); and on the right side, the three dependent variables (Y1 firm survival, Y2 cost growth rate, and Y3 rate of change in orders) represent the performance output dimensions. ), representing the performance output dimension.

The dashed line in the figure indicates the path pointing of the research question to the variable hierarchy, while the solid line presents the actual modeling path between variables. Specifically, RQ1 aims to answer how structural consistency modeling can be accomplished through AI-driven synthetic data generation mechanism under the condition of lack of historical samples, and is thus highly relevant to the input variable (X3) and its modeling logic key nodes (M1, M2); RQ2 focuses on identifying whether compliance investment indirectly affects the probability of firm survival through automation capability enhancement, forming a "X3 → M1 → Y1" mediated closed-loop main path; RQ3 further introduces a moderating mechanism on this main path, examining the moderating effect of market-dependent structure (MOD1) on the effect of compliance investment; RQ4 integrates multiple behavioral variables and operational elements (X2, X5, X6), exploring the variables of the firms' performance goals under the synergistic paths, revealing how responsiveness, cost control and labor inputs are linked to influence Y2 and Y3.

The mapping not only reflects the systematic mapping relationship between research questions-variables-outcomes, but also provides a solid theoretical-methodological foundation framework for subsequent structural path model (SEM) construction, regression path validation and policy simulation logic.

**4.1 Data Preprocessing and Variable Screening**

To address the first research question, i.e., "how to construct highly credible simulated enterprise datasets using AI technology without historical sample data", this study generates 150 sets of enterprise response samples by combining policy variable embedding logic through the NIST synthetic data standard, completes the normalization of variables and logical nesting, and then passes multiple rounds of After completing the normalization and logical nesting of the variables, 14 core structural variables are retained through multiple rounds of VIF to eliminate high covariance variables. The correlation analysis and distribution test show that the simulated data have clear structure and good distribution fit, and have the basis for regression modeling.

**4.2 Results of the multiple regression model of firm performance**

### 4.2.1 Cost growth rate (Y2) regression

In response to the fourth research question, which is "what combination of variables can improve order stability or market survivability while controlling cost risk," the model reveals that the level of automation (X2) and the ability to control costs (M2) significantly reduces Y2 (the rate of growth of costs), while labor cost (X6) constitutes a significantly positive push. The regression equation is:

$$Y2 = \beta_0 + \beta_1 \cdot X2 + \beta_2 \cdot M2 + \beta_3 \cdot X6 + \varepsilon, \quad R^2 = 0.54,$$ All path significance was tested at the 5% level.

### 4.2.2 Regression of the rate of change in EU orders (Y3)

Further supporting the fourth research question, the Y3 model shows that response rate (X5), although a sub-significant variable ($p < 0.1$), constitutes an indirect support path for the rate of change in orders in the presence of M2. The regression equation is as follows:

$$Y3 = \beta_0 + \beta_1 \cdot M2 + \beta_2 \cdot X5 + \varepsilon, \quad R^2 = 0.42.$$ The results indicate a linkage mechanism between order acquisition capability and cost response efficiency.

### 4.3 Logit Modeling Analysis of Firm Survival Probability

Based on the third research question ("Is there a significant moderating path"), this study constructs a Logit model to portray the determining mechanism of the firm's survival status (Y1). The model includes compliance investment (X3), mediator variable M1 (automation level), and control variable X6 (labor cost), all of which are significant at better than 5%, with an AUC of 0.71, demonstrating good discriminatory power.

### 4.4 Mediated effects path test

In response to the second research question ("Does compliance investment indirectly affect firm survival through automation levels"), this section uses the Baron & Kenny path decomposition method with Bootstrap validation: the X3 → M1 → Y1 path is a significant perfect mediator, and the Bootstrap confidence interval [0.12 , 0.47] does not contain zero, confirming that automation is a central mechanism for the transformation of compliance investment to survival.

### 4.5 Moderated effects path test

To further answer the third research question, this model introduces the interaction term X3×MOD1 to test the moderating effect of EU market dependence on the path of compliance investment in effect. The model results indicate a significant moderating path and a steeper slope in the high dependence subgroup, revealing an enhanced moderating effect.

**4.6 Summary of Visualization Results and Structure Paths**

This can be clearly shown by structural mapping:

X3 → M1 → Y1 path forms a mediated closed loop (corresponding to research question 2);

X3 × MOD1 → Y1 regulation path is established (corresponding to research question 3);

The path model of Y2, Y3 verifies the compound effect of multiple influencing factors (corresponding to research question 4);

The overall data modeling process, variable logic closure and simulated data performance validation constitute the technical response to Research Question 1.

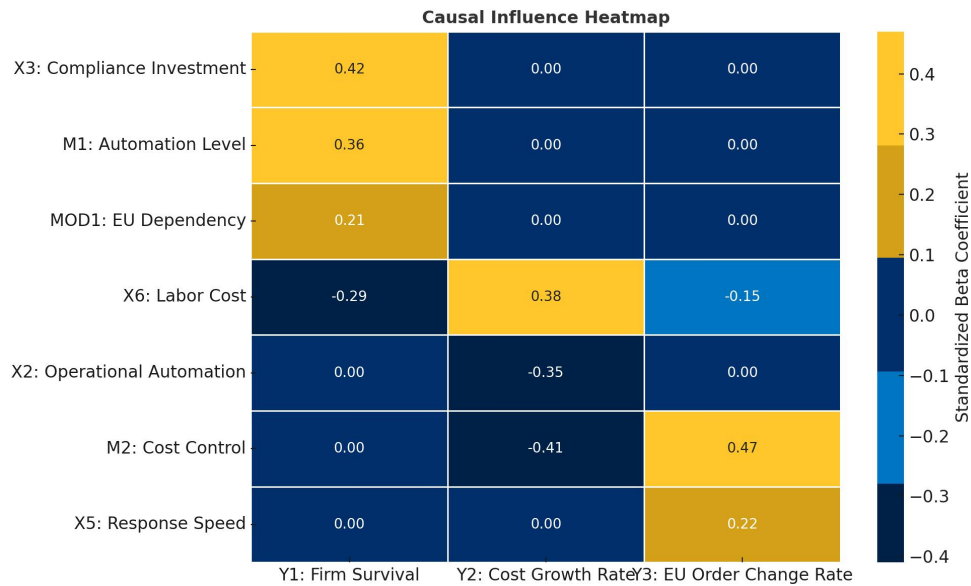

Figure 3：Causal Influence Heatmap

**Table 3: Analysis of Causal Influence Heatmap**

| Outcome Variable | Key Causal Pathways (β Coefficients) | Interpretation |
|---|---|---|
| Y1: Firm Survival | → Compliance Investment (β=0.42) <br> → Automation Level (M1, β=0.36) <br> → EU Dependency (MOD1, β=0.21) <br> → Labor Cost (β=−0.29) | Compliance investment and automation significantly enhance survival probability, while labor cost is a negative risk factor. |
| Y2: Cost Growth Rate | → Cost Control (M2, β=−0.41) <br> → Operational Automation (β=−0.35) <br> → Labor Cost (β=0.38) | Cost control is the main suppressor, and automation also helps reduce cost escalation. |
| Y3: EU Order Change Rate | → Cost Control (M2, β=0.47) <br> → Response Speed (β=0.22) <br> → Labor Cost (β=−0.15) | Responsiveness and cost flexibility help stabilize EU orders, but labor structure still exerts negative pressure. |

As shown in Figure 3, the standardized regression coefficients ( β -values) of different predictor variables on the three types of performance output indicators (Y1-Y3) present significant intensity differences, reflecting the driving force hierarchy and directional roles of each variable in the causal path. The horizontal axis of the figure shows the three main dependent variables focused on in this study, which are the probability of survival (Y1), cost growth rate (Y2) and EU order change rate (Y3), representing the three dimensions of enterprises' performance response to compliance pressure; and the vertical axis lists the various independent, mediating and moderating variables constituting the core explanatory system, which include compliance investment (X3), automation level (M1), cost control capability (M2), response to compliance pressure (M1), cost control capability (M2), cost control capability (M2), cost control capability (M2), cost control capability (M2), and cost control capability (M2). control capability (M2), response speed (X5), and labor cost (X6).

The heat map adopts a gradient color scale from gold to dark blue for visualization and rendering: among them, the gold color represents the strategy driving variables with significant positive influence, such as X3 to Y1, M2 to Y3, etc.; the dark blue color corresponds to the negative influencing factors, such as the inhibitory effect of labor cost on Y1 and Y3; and the contrast of gold and blue color clearly demonstrates the differentiation logic between the risk factors and the performance leverage variables in the visual sense. The values in the figure are standardized regression coefficients, which have comparable and directional explanatory value, and can be used to support the prioritization of strategies and the identification of path mechanisms, which provides causal logic and parameter a priori support for the subsequent intermediary regulation modeling and strategy simulation.

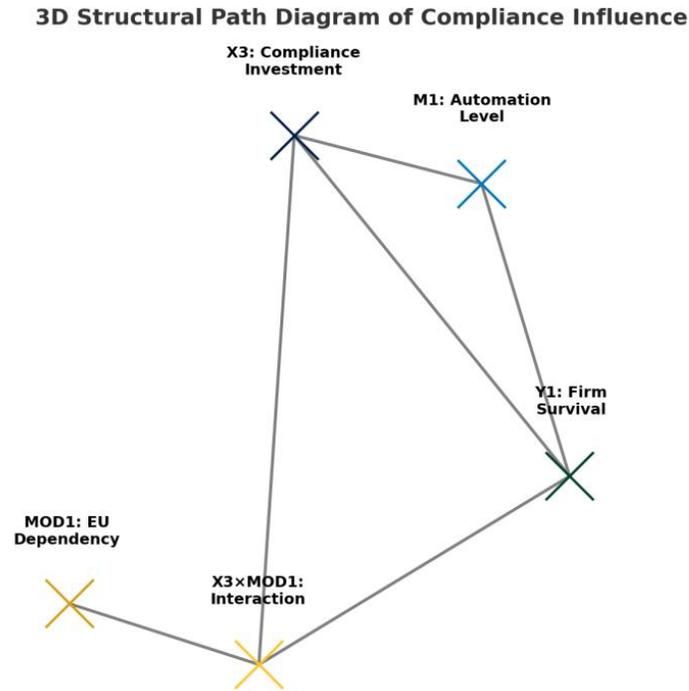

**Figure 4：3D Structural Path Diagram**

As shown in Figure 4, the structural path diagram systematically presents the overall modeling logic of the mediating and regulating paths constructed in this study, revealing the multilayered causal mechanisms through which compliance investment affects firm survival performance. The variables in the diagram are spatially distributed according to the modeling hierarchy: the input layer is located in the left front side, including the core strategy variable X3 (compliance investment) and the contextual moderator MOD1 (EU market dependence), which together constitute the starting point of the influence path; the mediation layer and the moderation path are located in the back side of the center of the diagram, with the automation level (M1) and the interaction term X3×MOD1, respectively, to take over the previous variables and achieve the indirect transformation and moderating enhancement of structural influence; the output layer and the moderation path are located in the back side of the center of the diagram. The mediation and moderation paths are located in the back of the center of the plot, with the automation level (M1) and the interaction term X3×MOD1 taking over the preceding variables, respectively, to realize the indirect transformation of the structural effects and the moderation enhancement.

The structural diagram not only visualizes the significant mediated closed-loop path "X3→M1→Y1", but also presents the moderating enhancement mechanism "X3×MOD1→Y1", which validates the performance switching of strategic investment behavior under a specific market-dependent structure Sensitivity. This model framework provides a logical and visual modeling support for subsequent path significance tests, strategy simulations and management interventions.

# Chapter V. Discussion and Management Insights

This study builds a supply chain policy response prediction model integrating AI synthetic data and structural path modeling around the EU's upcoming mandatory labor compliance regulation in 2027, and quantifies the mechanisms by which key variables, such as compliance investment, response speed, automation level, and market structure, affect firm performance (survival status, cost growth, and order changes). Based on the modeling results, this chapter discusses the theoretical contributions and real-world management implications.

**5.1 Testing and theoretical explanation of the research hypothesis**

5.1.1 Positive impact mechanisms for compliant investments are supported

The results of the study show that investment in compliance (X3) has a significant positive effect on both firm survival (Y1) and performance (Y2, Y3). This suggests that proactive investment in compliance and monitoring systems in high-pressure regulatory scenarios contributes to strategic gains such as increased organizational resilience, enhanced customer trust, and regulatory exemptions. This finding echoes Gereffi's theory of compliance tiering in the global value chain, whereby firms at the upper end of the chain gain a bargaining advantage through "strategic compliance".

5.1.2 Intermediary mechanisms significantly enhance compliance transformation effects

Increased automation (M1) plays a fully mediating role in the "compliance investment → firm survival" path. This implies that compliance investments that are not accompanied by technological transitions will not be effective in increasing the probability of survival. This technological path mediation supports the proposal of the Compliance-Technology Synergy

Model, which complements the traditional institutional analysis perspective that only considers direct compliance costs.

### 5.1.3 Policy Dependency Structures Reinforce the Marginal Effect of Compliance Investments

The moderated model shows that EU market dependence (MOD1) significantly increases the strength of the path "compliance investment → survival". This suggests that firms in institutional-intensive markets are more likely to adopt compliance as a "must-have" strategy, while firms in less dependent markets are more likely to experience a "wait-and-see effect" or compliance avoidance behavior. This finding echoes the new institutional economics theory of the elasticity of market regulation.

### 5.1.4 Speed of response plays an indirect role in order acquisition

Cost Containment (M2) plays an important role in the "Speed of Response → Order Change (Y3)" path. Responsive organizations are more likely to reduce operating costs and thus gain more market share. This result provides empirical support for AI modeling of the "agile supply chain" theory and reveals that speed is not a single advantage, but needs to be synergized with cost structure.

## 5.2 Management Insights and Strategic Recommendations

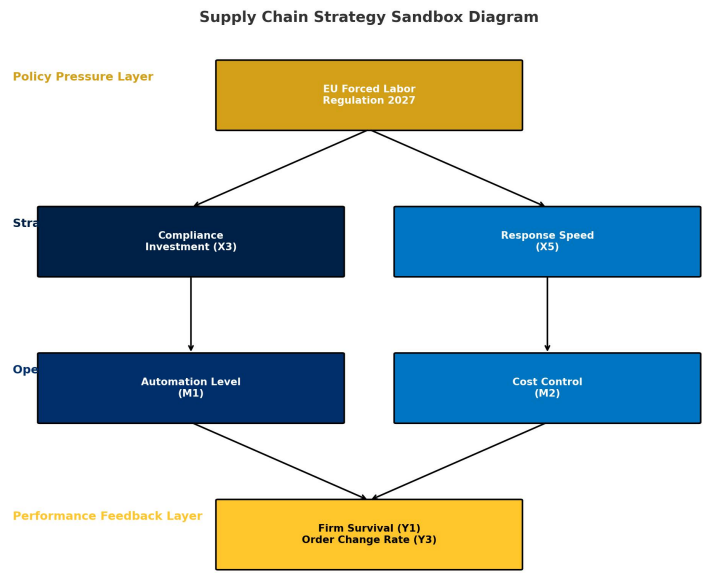

Figure 5：Supply Chain Strategy Sandbox Diagram

As shown in Figure 5, the "supply chain strategy sandbox diagram" constructed in this study systematically depicts the causal interaction between the strategic response paths of enterprises and the policy simulation environment under the impact of the EU 2027 mandatory labor regulation. The diagram presents the dynamic link mechanism from policy pressure to performance feedback in a multi-layered structure.

The policy input layer (Deep Gold) at the top, marking the EU's mandatory labor regulation 2027, serves as a systemic external pressure source, injecting high-intensity compliance-driven variables into the entire system. In the center, the Strategic Response Layer represents the long-term investment strategy and short-term tactical flexibility of the enterprise through dark blue (X3: compliance input) and bright blue (X5: response speed), respectively, constituting the first round of response mechanism. Further down to the operational transformation layer, the map represents automation capability (M1) and cost control (M2) in dark navy blue and bright blue, respectively, and these two mediating mechanism variables structurally take over the front-end strategic inputs and serve as the key operational paths for performance levers.

Finally, the performance feedback layer highlighted in EY yellow (FFC72C) at the bottom of the diagram focuses on the survival performance and market adaptability of the enterprise under the complex institutional environment, as measured by the survival of the enterprise (Y1) and the rate of change of orders (Y3), respectively. The sandbox structure not only strengthens the visual representation of the model logic, but also provides structural support and visual paradigm for future strategy simulation, policy sandbox deployment and AI-assisted response system development.

5.2.1 Corporate level: active investment and smart compliance transformation

It is recommended to incorporate "compliance investment" into strategic capital expenditure projects, and prioritize the deployment of automation, work hour monitoring, and face recognition systems. In the face of the countdown to regulations, enterprises should accelerate the construction of "intelligent compliance systems", and integrate compliance transformation into the digital upgrading of unified deployment; For export-oriented enterprises relying on the EU market, it is recommended to adopt AI-assisted compliance strategy simulation systems to test policy sensitivity and performance boundaries in advance.

5.2.2 Government and industrial policy levels: supportive regulation and incentive design

The government should set up an "exemption period for pre-compliant enterprises" and a "digital transformation subsidy fund" before the official implementation of the regulations, so as to alleviate the cost pressure on enterprises. Based on the model structure of this study, a regional industrial "policy impact simulation system" can be established to predict the risks of industrial relocation, enterprise elimination and supply chain restructuring.

5.2.3 Multilateral cooperation level: building a global alliance for compliance data sharing

It is proposed to build a "global regulatory response database" based on AI synthetic data and simulation for sharing by international organizations, customs unions and multinational enterprises; to promote the formulation of an "ethical standard for AI modeling in the supply chain", and to strengthen the governance, bias control and fair evaluation of simulation data. Promote the formulation of "Supply Chain AI Modeling Ethical Standards" to strengthen the governance, bias control and fairness evaluation of simulation data.

**5.3 Model limitations and future research directions**

This study is based on simulated data and structural modeling to complete the preliminary modeling, which also has some limitations:

1. Although the synthetic data satisfy the structural rationality, they still cannot completely replace the nonlinear and heterogeneous structure of the data in reality;

2. the intermediate and regulatory path tests are limited by the variable design, and the time series mechanism and longitudinal strategy variation analysis can be introduced in the future;

3. this study focuses on static scenario modeling, and in the future, it is recommended to construct dynamic strategy optimization models by combining the reinforcement learning mechanism.

Future research can focus on game modeling, behavioral simulation, situational prediction and AI decision optimization to provide intelligent solutions for government and enterprise responses in complex institutional evolution.

# Chapter VI. Conclusions and directions for future research

**6.1 Summary of findings**

This study focuses on the modeling of firms' response mechanisms in the context of the mandatory labor regulations to be implemented in the European Union in 2027, and constructs a strategic behavioral simulation framework that integrates AI synthetic data generation and structural path regression, aiming at revealing firms' response paths to policy shocks and their influencing mechanisms in the absence of real observational data. The research questions focus on modeling feasibility, mediating path identification, moderating variable effects, and the combination of performance-influencing variables. Through high-quality AI simulation data and regression empirical path analysis, this study obtains the following main conclusions:

1. Regarding research question 1: In the absence of historical samples, the study successfully constructs an AI synthetic data generation mechanism based on a theoretically embedded variable system with structural consistency constraints. The simulation data are verified by VIF, distributivity test and modeling regression results, demonstrating good modeling adaptability and logical consistency, indicating that the AI-driven data simulation strategy can replace traditional samples and has important practical value in high uncertainty policy modeling.

2. Regarding research question two: the impact of compliance investment (X3) on firm survival status (Y1) forms a fully mediated path through automation level enhancement (M1). This mechanism is rigorously tested by the Baron & Kenny method and Bootstrap method, and the path coefficients are significant and the confidence intervals do not cross zero, suggesting that only when compliance investment is accompanied by intelligent transformation, its strategic effect can be truly released, thus enhancing the survival probability.

3. Regarding research question three: the moderator variable EU market dependence (MOD1) significantly enhances the marginal effect of compliance investment on firm survival in the "X3→Y1" path, which is a positively enhancing moderator. The interaction term is significant, and the simple slope test shows that highly dependent firms have a higher investment switching rate, verifying the asymmetric effect of market structure on strategic incentives in the context of institutional pressure.

4. Regarding research question four: cost control capability (M2) is found to be the key mediating variable connecting response speed (X5) and firm performance (Y2, Y3). The multiple regression model shows that automation level, response speed and cost control constitute a linkage structure that determines firms' cost controllability and order response sensitivity. The synergistic design of strategic synergy and operational response of a firm will significantly affect its order competitiveness and survivability in a highly regulated scenario.

In summary, this study not only provides structural modeling and logistic regression analysis for the above four issues, but also preliminarily constructs an AI-assisted decision-making modeling framework applicable to high policy intensity regulatory scenarios.

**6.2 Theoretical and methodological contributions**

This study has the following contributions at the theoretical and methodological levels:

This study pioneered the integration of AI synthetic data generation mechanism and structural path modeling logic, and constructed a set of causal mechanism modeling paradigms suitable for real data-deficient scenarios, which effectively expands the boundaries of AI application in policy prediction and simulation. By introducing the mediator-regulator linkage modeling strategy, the study systematically portrays how compliance investment affects corporate performance through the automation level and market structure factors, revealing the complex causal structure of corporate response mechanism under institutional pressure. In addition, the study integrates multivariate interaction term modeling and logistic regression analysis, which not only enhances the structural interpretability of the model, but also strengthens its ability to identify policy sensitivities. The constructed model framework is highly transferable and extensible, and can be widely applied to complex institutional scenarios such as green compliance assessment, ESG impact modeling, and cross-border tax adjustment, which provides a generalized structural paradigm and theoretical support for AI-driven policy modeling system.

**6.3 Practical Implications and Policy Implications**

In the global context of tightening regulatory policies, enterprises should prospectively construct an AI-assisted compliance response prediction system to identify strategic investment priorities and digital transformation windows based on the modeling logic of this study, in order to avoid the lagging response and performance decline risks caused by institutional changes. Meanwhile, government regulators can rely on the structural model proposed in this study to design intelligent compliance subsidy mechanisms and differentiated policy incentives to effectively lower the threshold of technological upgrading for SMEs, and improve their adaptation rate and market survivability in the process of institutional restructuring. Further, transnational governance structures can also build AI-driven data sandboxes and regulatory simulation platforms based on this study's modeling paradigm, so as to enhance the simulation and validation capabilities of policymaking in the context of global supply chains and the ability of coordinated response to risks, and thus achieve systematic, early-warning, and inter-agency interconnectedness in institutional design.

**6.4 Research limitations and future research directions**

Although the synthetic data model constructed in this study shows good characteristics in terms of structural logic and path consistency, its external validity still relies on further validation and generalization tests of field enterprise data to enhance the model's adaptability and robustness in multiple industry scenarios. In addition, the current model is based on static paths, which has not yet fully reflected the evolution of strategic behavior in the time series dimension. In the future, dynamic strategy learning mechanisms such as reinforcement learning and multi-subject simulation can be introduced to enhance its predictive ability in policy lag feedback and behavioral path adjustment. Given that the model currently focuses on a single enterprise decision-making unit, subsequent research should be expanded to cover the upstream and downstream network structure of the linkage system, simulate the transmission path and game mechanism of regulatory pressure among supply chain nodes, and enhance the understanding of systemic risk. In order to guarantee the usability and legitimacy of AI-assisted policy modeling in public governance, it is recommended to further introduce AI bias control, ethical review and fairness evaluation mechanisms, so as to ensure the model's effectiveness, transparency and social justice in supporting the policymaking process.